\documentclass[aps,prb,twocolumn,showpacs,preprintnumbers,amsmath,amssymb,superscriptaddress]{revtex4-2}
\usepackage{makecell}
\usepackage{float}
\usepackage{graphicx}
\usepackage{dcolumn}
\usepackage{bm}
\usepackage{tabu}
\usepackage{array}
\usepackage{bm}
\usepackage{amssymb}

\usepackage{comment}

\usepackage{graphicx}
\usepackage{amsthm}
\usepackage{txfonts}
\usepackage{dcolumn}
\usepackage{bm}
\usepackage{hyperref}
\hypersetup{colorlinks,citecolor=red,filecolor=black,linkcolor=blue,urlcolor=blue}
\usepackage{lipsum}
\usepackage{subcaption}
\usepackage{subfloat}
\usepackage{caption}
\usepackage{natbib}

\usepackage[utf8]{inputenc}
\usepackage{calc}
\usepackage{accents}

\def\be{\begin{eqnarray}}
\def\ee{\end{eqnarray}}
\usepackage{stackengine}

\begin{document}
\title{Engineering Kekul\`{e} Superconductivity from Layer-selective Interactions in Rhombohedral Multi-layer Graphene}

\author{Hung Dinh Nguyen}
\affiliation{Department of Physics, University of Nevada, Reno, Nevada 89557, USA}
\author{Yafis Barlas}
\affiliation{Department of Physics, University of Nevada, Reno, Nevada 89557, USA}

\begin{abstract}
At weak coupling, finite-momentum superconductivity is typically associated with broken time-reversal or inversion symmetry of the Fermi surface. Here, we show that lattice-scale pair-density-wave order in rhombohedral multilayer graphene can arise from layer/orbital-dependent pairing interactions, band chirality, and Dirac-point-centered Fermi surface topology while preserving both symmetries. Using mean-field theory and comparing finite momentum sectors ${\bm Q}= \pm 2{\bm K}_{D}$ with the ${\bm Q}= 0$ superconducting state, we find that layer-dependent interactions of opposite signs ($V_{1A}=-V_{JB}=-|V|$) favor an intra-valley Kekul\`{e} state with center-of-mass momentum (${\bm Q}=\pm 2{\bm K}_D$). In the presence of a time-reversal and inversion symmetry-preserving Kane–Mele mass ($\lambda$) this state appears only above a critical carrier density ($n^{\mathrm{crit}}_{K}(\lambda,J)$). The two superconducting condensates exhibit opposite chirality, $J(-J)$ for ${\bm K}_D(-{\bm K}_D) $ valleys, thereby preserving time-reversal and inversion symmetry. We map the phase diagram and analyze the dependence of $T_c$ on the chirality index $J$ and $\lambda$. We also evaluate the superfluid stiffness in the Kekul\`{e} superconducting state, thereby determining the Berezinskii-Kosterlitz-Thouless (BKT) transition temperature. Our results show that orbital-dependent interactions in the presence of band chirality favor finite-momentum pairing in time-reversal and inversion symmetric Dirac materials.
\end{abstract}
\maketitle

\section{Introduction}

Due to perfect nesting of a time-reversal- and inversion-symmetric Fermi surface, existing routes to finite-momentum superconductivity generally rely either on symmetry breaking or strong coupling. For example, Zeeman splitting creates concentric spin-resolved Fermi surfaces at different Fermi wavevectors that can favor the Fulde-Ferrell-Larkin-Ovchinnikov (FFLO) state~\cite{PhysRev.135.A550,LOpaper}, but its realization is restricted to a narrow class of materials in which orbital effects are sufficiently weak~\cite{doi:10.1143/JPSJ.76.051005}. Likewise, helical superconductivity~\cite{PhysRevLett.87.037004,PhysRevB.76.014522,HelicalSCFu,lqry-yq57} requires broken spatial symmetries and has remained elusive. Even time reversal and inversion symmetry preserving pair-density-wave (PDW) states in Cuprate superconductors~\cite{RevModPhys.87.457,PDWreview}, typically arise from strong coupling, competing ordered phases or fluctuations~\cite{PhysRevLett.88.117001,PhysRevB.81.020511,PhysRevX.4.031017,PhysRevB.89.165126,RevModPhys.87.457,PDWreview,PhysRevLett.125.167001,PDWDavis,PhysRevLett.99.127003,Agterberg2008,Berg2009,PhysRevB.91.104512,PhysRevB.97.174510,PhysRevB.97.174511} and often coexist with uniform superconductivity, obscuring both its microscopic origin and experimental signatures~\cite{vortexhalosPDW,PhysRevLett.102.207004,PhysRevB.77.174502,QOcupratesPDW}. Therefore, realizing finite-momentum pairing as a genuine weak-coupling instability in a time-reversal- and inversion-symmetric parent state would be particularly compelling. 

Rhombohedral multilayer graphene (RMG) provides a natural setting for lattice-scale finite-momentum superconductivity. Its chiral bands~\cite{PhysRevB.77.155416,10.1143/PTPS.176.227,PhysRevB.80.165409,PhysRevB.82.035409,Lui2011} and Dirac-centered Fermi surfaces host a rich landscape of correlated and topological phases~\cite{Bao2011,Zhang2011,10.1021/nl303375a,Shi2020,PhysRevLett.127.187001,Zhang2023,Liu2024,Yang2025,kumar2025superconductivitydualsurfacecarriersrhombohedral,Zhou2021SC,Han2025,qin2026stripeordermetallicsuperconducting,Zhou2021Half,nguyen2025hierarchysuperconductivitytopologicalcharge,Geier2025,yang2024topologicalincommensuratefuldeferrelllarkinovchinnikovsuperconductor,PhysRevB.111.174523,wy3f-hgr9,fcdc-9lm3,zfmh-rjzc,fdz1-dbf6,gt8h-czf3,Qin_2026,PhysRevB.111.014508}, including spin- and valley-polarized metals and quantum anomalous Hall states~\cite{Geisenhof2021,Han2024AH,Lu2024AH,doi:10.1126/science.adk9749,PhysRevLett.133.206503,Winterer2024AH,doi:10.1126/science.adj8272,PhysRevB.110.205130,Lu2025AH}. A particularly simple route to finite-momentum pairing arises in the spin- and valley-polarized quarter-metal~\cite{Geier2025,yang2024topologicalincommensuratefuldeferrelllarkinovchinnikovsuperconductor,PhysRevB.111.174523,wy3f-hgr9,fcdc-9lm3,zfmh-rjzc,fdz1-dbf6,gt8h-czf3,Qin_2026,PhysRevB.111.014508}, where a single Fermi surface centered at ${\bm K}_D$ enforces intra-valley pairing with center-of-mass momentum ${\bm Q}=2{\bm K}_D$, producing a Fulde--Ferrell-like Kekul\`{e} PDW~\cite{wy3f-hgr9,fcdc-9lm3,zfmh-rjzc}. 
 
 When both valleys are occupied, however, as in half-metallic or spin- and valley-unpolarized states, finite-momentum superconductivity requires intra-valley pairing to overcome the competing zero-momentum inter-valley channel. This can occur in a spin-polarized Haldane phase, where Chern-band topology favors intra-valley pairing and stabilizes a lattice-scale PDW even at weak coupling~\cite{barlas2025quantumgeometryinducedkekule}. The resulting chiral superconducting state exhibits a characteristic Kekul\'{e} modulation in the superconducting gap~\cite{PhysRevB.82.035429,PhysRevB.93.155149,barlas2025quantumgeometryinducedkekule}. 

In this paper, we show that a lattice-scale PDW can emerge as a weak-coupling instability of the spin- and valley-unpolarized metallic state of RMG, driven by the interplay of layer-selective interactions and band chirality. Specifically, we consider the case in which the effective interaction is attractive in one layer and repulsive in the other. To keep the analysis transparent, we restrict ourselves to onsite interactions, which select the opposite-spin pairing sector. Using a generalized chiral model for RMG~\cite{PhysRevB.77.155416}, we project layer-dependent onsite interactions onto the low-energy conduction band and show that the pairing problem separates into a conventional inter-valley channel with ${\bm Q}=0$ and intra-valley Kekul\`{e} channels with ${\bm Q}=\pm 2{\bm K}_{D}$. We find that a layer-symmetric attraction, $V_{1A}=V_{JB}<0$, favors the conventional zero-momentum state, whereas an interaction with equal magnitude and opposite signs on the two outer-layer sublattices, $V_{1A}=-V_{JB}$, suppresses the inter-valley pairing channel and makes the finite-momentum intra-valley pairing state the leading instability. The corresponding intra-valley superconducting order inherits the chiral phase winding of the Bloch wavefunctions and produces a fully gapped state with a tripled unit cell and Kekul\`{e} superconducting pattern~\cite{PhysRevB.82.035429,PhysRevB.93.155149,barlas2025quantumgeometryinducedkekule}. This state consists of two superconducting condensates with opposite chirality, $J(-J)$ for ${\bm K}_D(-{\bm K}_D) $ valleys, thereby preserving time-reversal and inversion symmetry.

In the presence of a finite Kane-Mele mass ~\cite{PhysRevLett.95.226801} $\lambda \neq 0$, and for purely anti-symmetric interactions, the competition between the two pairing channels is determined by the dimensionless ratio $X=\lambda/\mu$, where $\mu$ is the chemical potential. Our mean-field analysis shows that the Kekul\'{e} state becomes dominant for $X<\sqrt{2}-1$. This yields a superconducting phase boundary independent of the overall interaction strength, but dependent on the electron density $n_{2D}$, band chirality $J$, and $\lambda$. The intra-valley Kekul\`{e} superconductor requires $n_{2D} \geq n^{\mathrm{crit}}_{K}(\lambda,J)$. We also calculate the transition-temperature scaling as a function of carrier density, Kane-Mele mass, chirality index, and interaction strength. For representative weak-coupling parameters, the predicted mean-field transition temperatures are $T_{\mathrm{c}} \sim 10 \mathrm{K} - 100 \mathrm{mK}$ (depending on the interaction strength).

To determine whether the finite-momentum condensate possesses sufficient phase rigidity, we calculate the superfluid stiffness. We show that in the weak-coupling limit, the geometric contribution to the superfluid stiffness is small compared to the conventional contribution and scales as $D_{s} \sim \mu$ throughout the Kekul\'{e} regime, ensuring a finite and positive superfluid stiffness. The resulting Berezinskii-Kosterlitz-Thouless (BKT) transition temperature ($T_{BKT}$) exhibits a density-dependent suppression that depends on the chirality index $J$. The maximum reduction corresponds to the ratio of ($T_{BKT}/T_{\mathrm{c}} \sim 0.8$). The experimental signature for the Kekul\`{e} modulation and the optical signatures of the valley-locked chiral condensates are also discussed.

The remainder of the paper is organized as follows: In Sec.~\ref{Section: Model}, we introduce the chiral Kane-Mele model and the sublattice-dependent onsite interactions and determine the band-projected pairing interactions for inter-valley and intra-valley pairing sectors. In Sec.~\ref{Section: Mean field}, we formulate the mean-field gap equation in the different COM momentum sectors. In Sec.~\ref{Section: Kekule SC}, we analyze the competition between conventional and Kekul\'e superconductivity and derive a simple criterion for when the Kekul\'e channel becomes dominant. In Sec.~\ref{Section: SC phase}, we discuss the superconducting phase diagram and the dependence of the transition temperature on the chirality $J$, spin-orbit coupling strength $\lambda$, carrier density $n_{2D}$, and interaction strength $|V_z|$. In Sec.~\ref{Section: Superfluid stiffness}, we compute the superfluid stiffness of the Kekul\'e state and use the Nelson-Kosterlitz criterion to estimate the BKT transition temperature. In the final section, Sec.~\ref{Section: Discussion}, we discuss the possible experimental signatures of the proposed Kekul\'{e} superconductivity in RMG stacks, remote hopping effects, and conclude our findings.

\section{Model}
\label{Section: Model}
 
In the presence of a Kane-Mele spin-orbit coupling (SOC), the low-energy properties ($\epsilon \ll \gamma_1$) of $J$-layer rhombohedral graphene (RMG) systems can be described by an extended chiral Kane-Mele Hamiltonian \cite{PhysRevB.77.155416,PhysRevLett.95.226801}, 
\begin{equation} \label{eq:chiral_hamiltonian}
    H_0 =\xi_J p^J
    \left[
    \cos(J\phi_{\bm p})\hat{\sigma}_x
    + \tau_z \sin(J\phi_{\bm p})\hat{\sigma}_y
    \right]
    + \lambda \tau_z s_z\hat{\sigma}_z ,
\end{equation}
which acts on the two-component spinor $(\psi_{1A,\tau,s}({\bm p}),\; \tau \psi_{JB,\tau,s}({\bm p}))^T$, where $\hat{\sigma}$ acts in the sublattice/layer $(1A,JB)$ space, with sublattice $1A$ on the top layer and sublattice $JB$ on the bottom layer, $\tau=\pm 1$ labels the two valleys, and $s=\pm 1$ labels spin, and the parameter $\lambda$ is the Kane-Mele mass. The momentum ${\bm p}$ is measured relative to each valley center, with magnitude $p$ and polar angle $\phi_{\bm p}$, $J$ denotes the chirality index; for the minimal model of rhombohedral multilayer graphene (RMG), it is equal to the number of layers. The coefficient $\xi_J$ is determined by microscopic band parameters with $\xi_J=(\hbar v_F)^J/\gamma_1^{J-1}$, where $\gamma_1 \sim 400$ meV denotes the inter-layer hopping between Carbon atoms in adjacent layers that are stacked on top of each other, and $v_F$ denotes the Fermi velocity. We ignore remote hopping effects, which can lead to trigonal warping of the band structure, for now and address them later.

The Hamiltonian in Eq.~\eqref{eq:chiral_hamiltonian} gives a particle-hole symmetric gapped spectrum, $\epsilon_{\tau,s}({\bm p}) = \pm \sqrt{\xi_J^2p^{2J}+\lambda^2}$ with a gap $2 \lambda$. In our calculations, we primarily focus on the conduction bands. A convenient form of the conduction-band spinor is
\begin{equation} \label{eq:chiral_spinor}
    |u_{\tau,s}^{(+)}({\bm p})\rangle =
    \begin{pmatrix}
        e^{-i\tau J\phi_{\bm p}/2}\cos (\theta_{\tau,s}({\bm p})/2) \\
        e^{+i\tau J\phi_{\bm p}/2}\sin [\theta_{\tau,s}({\bm p})/2]
    \end{pmatrix},
\end{equation}
where $ \cos[\theta_{\tau,s}({\bm p})]=\lambda \tau s/\epsilon_{\tau,s}({\bm p})$ and $\sin[\theta_{\tau,s}({\bm p})]
= \xi_J p^J/\epsilon_{\tau,s}({\bm p})$. 
The angle $\theta_{\tau,s}({\bm p})$ controls the relative weight of the two sublattices, while the phases $e^{\pm i\tau J\phi_{\bm p}/2}$ encode the chiral winding of the Bloch wave function. After projection onto the conduction band, these two features determine the momentum- and valley-dependence of the pairing form factors. Since $H_0$ commutes with $s_z$, spin is conserved, and each electron on a valley Fermi surface has a well-defined spin. 

We consider sub-lattice (or layer) dependent on-site interactions which can be expressed as, 
\begin{equation} \label{eq:onsite_interaction}
H_{int} = \sum_{i,\alpha} V_\alpha n_{i \alpha \uparrow} n_{i \alpha \downarrow},
\end{equation}
where $n_{i\alpha s}=c_{i\alpha s}^\dagger c_{i\alpha s}$, $s=\uparrow,\downarrow$, and $\alpha=(1A,JB)$ labels the two low-energy layer/sublattice degrees of freedom of RMG. The couplings $V_{1A}$ and $V_{JB}$ describe onsite pairing interactions on these two sublattices. In this paper, we assume that at least one interaction channel contains an attractive component. 

Several microscopic mechanisms may generate such an interaction. This class of interactions can arise microscopically from electron–phonon-mediated pairing in the presence of Coulomb repulsion, due to substrate effects. A more directly controllable realization can be obtained through the superconducting proximity effect. For example, RMG may be placed on a conventional superconducting substrate, with only one outer layer directly coupled to the substrate. Although inter-layer hopping subsequently transfers superconducting correlations to the other graphene layers, the microscopic pairing source remains strongly layer asymmetric. This provides a natural way to realize a model in which an attractive interaction acts primarily on a selected layer.

\begin{figure}[t!]
    \centering
    \includegraphics[width=0.95\linewidth]{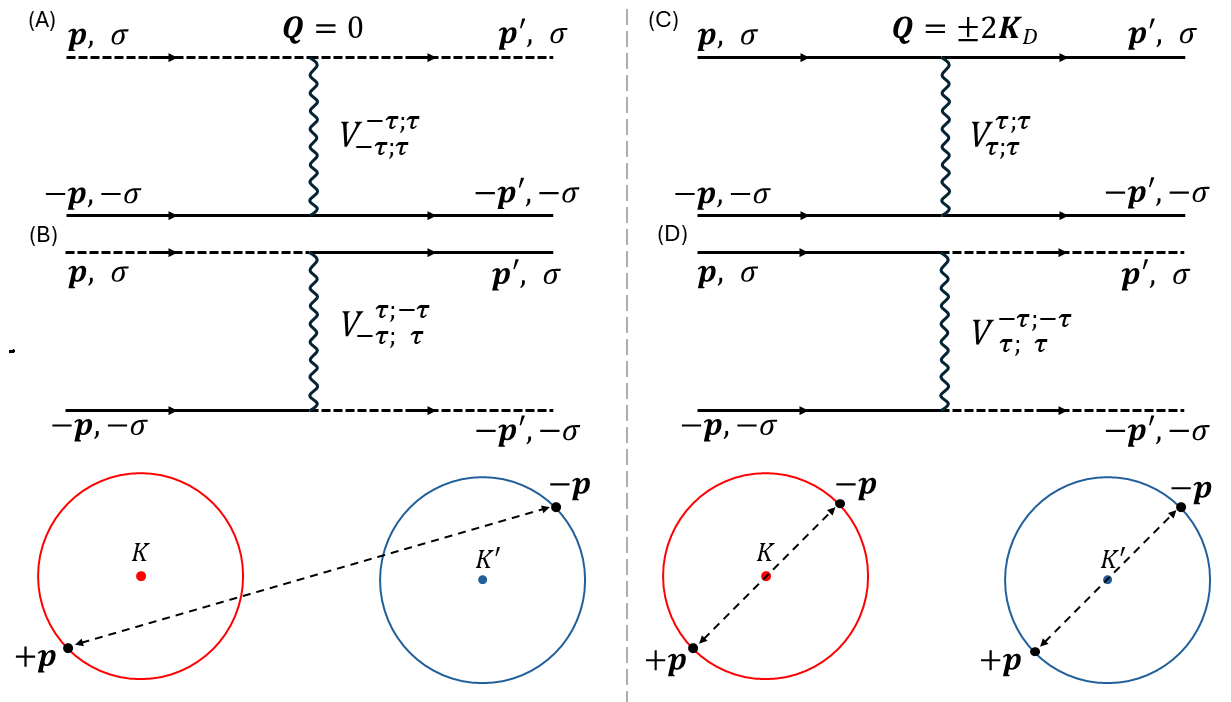}
    \caption{
    Schematic illustration of valley-resolved Cooper-pair scattering processes generated by the projected interaction.
    (A)-(B) In the ${\bm Q}=0$ sector, the two electrons originate from opposite valleys, so the pair carries zero COMM.
    (C)-(D) In the ${\bm Q}=\pm 2{\bm K}_D$ sector, both electrons originate from the same valley, so the pair carries finite COMM.
    In each sector, the projected interaction includes both direct pair scattering within a fixed valley configuration and channel-mixing processes that alter the pair's valley content.}
    \label{fig:pair_scattering_schematic}
\end{figure}

To proceed with the low-energy effective interaction, we project Eq.~\eqref{eq:onsite_interaction} onto the conduction band near the valley Fermi surface. After projection, we only keep interactions corresponding to the pair center-of-mass momenta ${\bm Q} =0$ and $ {\bm Q} = \pm 2 {\bm K}_D$. In general, Cooper pairs may carry an arbitrary center-of-mass momentum (COMM) ($\bm Q$). However, considerations based on Fermi-surface nesting restrict the dominant pairing instabilities to two possibilities, namely an inter-valley state with ${\bm Q} =0$ or a finite-momentum intra-valley state with $\bm Q = \pm 2 {\bm K}_D$. 

The projected interaction for ${\bm Q} =0$ can be expressed as
\begin{equation}
\label{eq:projected_interactionQz}
H_{\mathrm{int}}({\bm Q}=0)
=
\frac{1}{L^2}
\sum_{{\bm p},{\bm p'}}
\sum_{\tau,\tau'}
V_{\tau',-\tau'}^{\tau,-\tau}({\bm p},{\bm p'})
\,P^\dagger_{\tau,-\tau}({\bm p})\,
P_{\tau',-\tau'}({\bm p'}) , 
\end{equation}
where $P^{\dagger}_{\tau_1,\tau_2}({\bm p}) = a^{\dagger}_{\tau_2\downarrow}(-{\bm p})\, a^{\dagger}_{\tau_1\uparrow}({\bm p})$ denotes the Cooper pair operators, and $a^{\dagger}_{\tau s}({\bm p})$ creating a conduction-band electron in valley $\tau$ and spin $s$. The projected interaction for ${\bm Q} =\pm 2 {\bm K}_D$ can be expressed as
\begin{equation}
\label{eq:projected_interaction2KD}
H_{\mathrm{int}}({\bm Q}=\pm 2 {\bm K}_D)
=
\frac{1}{L^2}
\sum_{{\bm p},{\bm p'}}
\sum_{\tau,\tau'}
V_{\tau',\tau'}^{\tau,\tau}({\bm p},{\bm p'})
\,P^\dagger_{\tau,\tau}({\bm p})\,
P_{\tau',\tau'}({\bm p'}).
\end{equation}
In both Eqs.~\ref{eq:projected_interactionQz} and ~\ref{eq:projected_interaction2KD}, the projected matrix element is defined as 
\begin{equation}
\label{eq:projectedV}
V^{\tau_4,\tau_3}_{\tau_1,\tau_2}({\bm p},{\bm p}')
=\sum_{\alpha=1A,JB}
V_\alpha\,
u^{*}_{\alpha,\tau_4\uparrow}({\bm p})\,
u^{*}_{\alpha,\tau_3\downarrow}(-{\bm p})\,
u_{\alpha,\tau_2\downarrow}(-{\bm p}')\,
u_{\alpha,\tau_1\uparrow}({\bm p}').
\end{equation}
Equation~\eqref{eq:projectedV} shows how a local interaction in the orbital basis attains a momentum and valley dependence after projection to the conduction band.

The effective pairing interactions are represented by the four tree-level Feynman diagrams shown in Fig.~\ref{fig:pair_scattering_schematic}. Two of these processes contribute to the inter-valley pairing channel, whereas the remaining two contribute to the intra-valley pairing channel.  The inter-valley scattering, see Fig.~\ref{fig:pair_scattering_schematic} a) is denoted by $(V^{+-}_{+-}, V^{-+}_{-+})$, while the inter-valley exchange interaction, see Fig.~\ref{fig:pair_scattering_schematic} b) is denoted by $(V^{+-}_{-+}, V^{-+}_{+-})$ resulting in a conventional superconductor with zero COMM pairing. Same-valley pairs ${\bm Q}=\pm 2{\bm K}_D$ correspond to finite COMM Kekul\'e pairing. The interaction matrix element corresponding to the Kekul\'e pairing consists of the intra-valley pair scattering, see Fig.~\ref{fig:pair_scattering_schematic} c) denoted by $(V^{++}_{++}, V^{--}_{--})$, as well as Cooper pair tunneling corresponding to the interaction matrix elements $(V^{++}_{--}, V^{--}_{++})$ indicated by the Feynman diagram in Fig.~\ref{fig:pair_scattering_schematic} d). The projected interaction matrix elements for the chiral bands of the $J$-layer RMG for both ${\bm Q} =0$ and ${\bm Q} = \pm 2 {\bm K}_{D}$ are presented in Appendix ~\ref{app:AppendixMatelements}.

In the general case, one may also allow for a residual pairing of COMM ${\bm q}$. In the inter-valley sector, this corresponds to ${\bm Q}={\bm q}$, while in the intra-valley sector it corresponds to ${\bm Q}=\pm 2{\bm K}_D+{\bm q}$. We restrict ourselves to lattice-commensurate PDW order (i.e., ${\bm q}=0$), namely ${\bm Q}=0$ and ${\bm Q}=\pm 2{\bm K}_D$. However, if the pockets are strongly warped or if the relevant symmetries are broken, finite-${\bm q}$ pairing can become competitive. In RMG, remote hopping effects can lead to significant trigonal warping of the band structure. This warping can give rise to coexisting or competing PDW states with ${\bm q}\neq 0$. We restrict our analysis to carrier densities above the trigonal-warping scale, where the Fermi surface is simply connected and approximately circular.

\section{Mean-field analysis}
\label{Section: Mean field}

Applying a standard mean-field decoupling to Eq.~\eqref{eq:projected_interactionQz} and Eq.~\eqref{eq:projected_interaction2KD}, we reduce the projected interaction to a quadratic Bogoliubov-de Gennes Hamiltonian. Since the interaction is block diagonal in the commensurate pair momentum sector, each ${\bm Q}$ sector can be treated independently. The self-consistent equation for the superconducting gap $\Delta_{\tau_1,\tau_2}({\bm p})$ follows from the mean-field decoupling of the projected interaction,
\begin{equation}
\label{eq:general_gap_def_sec}
\Delta_{\tau_4\tau_3}({\bm p})
=
\frac{1}{L^2}
\sum_{{\bm p}'}
\sum_{ \tau_1,\tau_2}^{'}
V^{\tau_4,\tau_3}_{\tau_1,\tau_2}({\bm p},{\bm p}')
\big\langle
a_{\tau_2\downarrow}(-{\bm p}')\,
a_{\tau_1\uparrow}({\bm p}')
\big\rangle,
\end{equation}
where the prime restricts the summation indices to the set of inter-valley and intra-valley interaction matrix elements indicated in Fig~\ref{fig:pair_scattering_schematic}. The quasiparticle spectrum takes the form
$E_{\tau,\tau'}({\bm p})=\sqrt{\xi({\bm p})^2+|\Delta_{\tau,\tau'}({\bm p})|^2}$, where $ \xi({\bm p})=\epsilon({\bm p})-\mu$ and the anomalous pair average can be expressed as,
\begin{equation}
\label{Pairing average}
\big\langle
a_{\tau'\downarrow}(-{\bm p})\,
a_{\tau\uparrow}({\bm p})\big\rangle
=-\frac{\Delta_{\tau,\tau'}({\bm p})}{2E_{\tau,\tau'}({\bm p})}
\tanh\!\left[
\frac{\beta E_{\tau,\tau'}({\bm p})}{2}
\right].
\end{equation}
The inter-valley superconducting gap $\Delta_{\tau,-\tau}$ at ${\bm Q} =0$, and the intra-valley superconducting gap $\Delta_{\tau,\tau}$ at ${\bm Q} = \pm 2 {\bm K}_D$ determine the relative strength of the superconducting state that dominates for a given set of interactions $(V_{1A},V_{JB})$.

We use the linearized gap equations to determine the largest value of the critical temperature $T_{\mathrm{c}}$ for each $\bm Q $ sector. Near the transition temperature $T_c$, the gap amplitude is small; therefore, one can linearize the gap equation by taking $E({\bm p})\approx |\xi({\bm p})|$. It is convenient to define the superconducting order parameters in vector notation,
\begin{align}
&\vec{\Delta}_{{\bm Q}={\bm 0}}({\bm p}) = (\Delta_{+-}({\bm p}),\; \Delta_{-+}({\bm p}))^T, \\ 
&\vec{\Delta}_{{\bm Q}=\pm 2{\bm K}_D}({\bm p}) = (\Delta_{--}({\bm p}),\; \Delta_{++}({\bm p}))^{T},
\end{align}
where $\Delta_{+-}(\Delta_{-+})$ correspond to inter-valley pairs and $\Delta_{++}(\Delta_{--})$ correspond to intra-valley pairs. The gap equations within each ${\bm Q}$ sector reduce to the linear integral eigenvalue problem, which can be expressed as,
\begin{equation}
\label{eq:linearized_gap_sec}
\vec{\Delta}_{\bm Q}({\bm p})
=
-\frac{1}{L^2}\sum_{{\bm p}'}
V_{\bm Q}({\bm p},{\bm p}')
\frac{\tanh\!\left[\beta \xi({\bm p}')/2\right]}{2\xi({\bm p}')}
\,\vec{\Delta}_{\bm Q}({\bm p}').
\end{equation}
where the projected interaction kernel $V_{\bm Q}$ can be expressed as, 
\begin{equation}
V_{{\bm Q}={\bm 0}}({\bm p},{\bm p}')
=\begin{pmatrix}
V^{+-}_{+-}({\bm p},{\bm p}') & V^{+-}_{-+}({\bm p},{\bm p}') \\
V^{-+}_{+-}({\bm p},{\bm p}') & V^{-+}_{-+}({\bm p},{\bm p}')
\end{pmatrix}, 
\end{equation} 
and 
\begin{equation}
V_{{\bm Q}=\pm 2{\bm K}_D}({\bm p},{\bm p}')
=
\begin{pmatrix}
V^{--}_{--}({\bm p},{\bm p}') & V^{--}_{++}({\bm p},{\bm p}') \\
V^{++}_{--}({\bm p},{\bm p}') & V^{++}_{++}({\bm p},{\bm p}')
\end{pmatrix}.
\end{equation} 
The superconducting instability is determined by the eigenstructure of the projected interaction kernel $V_{\bm Q}$. For the case of a circular Fermi surface of the RMG, the projected interaction matrix kernel for both zero momentum ${\bm Q}=0$ and finite momentum ${\bm Q} = \pm 2{\bm K}_{D}$ are supplied in Appendix~\ref{app:AppendixMatelements}. Within the weak-coupling approximation, we only consider the pairing attraction with a thin shell ($|\xi_{\bm p}'| \leq \omega_c$) near the Fermi surface, which allows us to write 
\begin{equation}
\frac{1}{L^2}\sum_{{\bm p}'} \longrightarrow N(\mu) \int_{-\omega_c}^{\omega_c} d \xi_{{\bm p}'} \int_0^{2 \pi} \frac{d \phi_{{\bm p}'}}{2 \pi},
\end{equation} 
where $N(\mu)$ is the density of states at the Fermi energy $\mu$.

\section{Kekule Superconductivity}
\label{Section: Kekule SC}

\subsection{Superconductivity for $\lambda= 0$}

In the absence of a band gap, the spin-resolved conduction-band spinors have equal weight on the two outer-layer sublattices of RMG. For the inter-valley sector, due to the chiral nature of the conduction bands, only the layer-symmetric interaction $V_0$ survives in the ${\bm Q} =0$ sector (see the Appendix~\ref{app:AppendixMatelements} for the interaction matrix elements). The interaction kernel has no momentum dependence; therefore, the superconducting gap exhibits an s-wave order parameter with $\vec{\Delta}_{{\bm Q}={\bm 0}}({\bm p})= \Delta\,(1,\;1)^T$. For $V_{1A} = V_{JB} = -|V_0|$, the linearized gap equation becomes
\begin{equation}
\label{eq:Linearized_gap_Q0_noSOC}
\begin{pmatrix}
\Delta_{+-} \\
\Delta_{-+}
\end{pmatrix}
= \frac{|V_0|\eta(T_c)}{2}\,
\begin{pmatrix}
1 & 1 \\
1 & 1
\end{pmatrix}
\begin{pmatrix}
\Delta_{+-} \\
\Delta_{-+}
\end{pmatrix},
\end{equation}
where $\eta(T_c) = N(\mu) \ln (1.14\,\omega_c/(k_B T_c))$ encodes the Cooper instability, and the critical temperature is determined from $|V_0|\,\eta(T_c)=1$. As anticipated, an inter-valley superconducting instability exists when $V_0<0$, resulting in a conventional BCS ${\bm Q} =0$ s-wave superconductor. 

In contrast, in the intra-valley sector, ${\bm Q}=\pm 2{\bm K}_D$, the interaction matrix elements contain nontrivial angular dependence (see the Appendix~\ref{app:AppendixMatelements} for the interaction matrix elements). A momentum-independent order parameter does not survive the angular integration in this sector. Instead, the order parameter inherits the chiral phase winding of the Bloch spinors. We therefore use the ansatz $\vec{\Delta}_{{\bm Q}=\pm 2{\bm K}_D}({\bm p}) = ( \Delta_{--} e^{i\kappa J\phi_{\bm p}} ,\;\Delta_{++} e^{-i\kappa J\phi_{\bm p}})^{T}$, where $ \kappa=\pm 1$. These two choices $\kappa=\pm1$ correspond to the two equivalent chiralities selected by which sublattice interaction is attractive ($\kappa =1$ corresponds to $V_{1A} <0$ and $V_{JB} >0$). 

Substitution into the linearized gap equation and angular integration gives
\begin{equation}
\label{eq:Linearized_gap_Q2KD_noSOC}
\begin{pmatrix}
\Delta_{--} \\
\Delta_{++}
\end{pmatrix}
=
-\frac{(V_0+\kappa V_z) \eta(T_c)}{4}
\begin{pmatrix}
1 & 1 \\
1 & 1
\end{pmatrix}
\begin{pmatrix}
\Delta_{--} \\
\Delta_{++}
\end{pmatrix}.
\end{equation}
The leading solution has a uniform amplitude but carries opposite chiral phase winding in each valley,
$ \vec{\Delta}_{{\bm Q}=\pm 2{\bm K}_D}({\bm p}) = \Delta ( e^{i\kappa J\phi_{\bm p}},\; e^{-i\kappa J\phi_{\bm p}} )^T $,
with the critical temperature determined by $ \left(V_0+\kappa V_z\right)\eta(T_c) = -2$. In the purely symmetric limit, $V_{1A}=V_{JB}=-|V_0|$, the finite-momentum channel satisfies $|V_0|\eta(T_c)=2$ and is therefore subleading to the conventional ${\bm Q}= 0$ state, whose effective coupling is twice as large.

\begin{figure}
    \centering
    \includegraphics[width=0.95\linewidth]{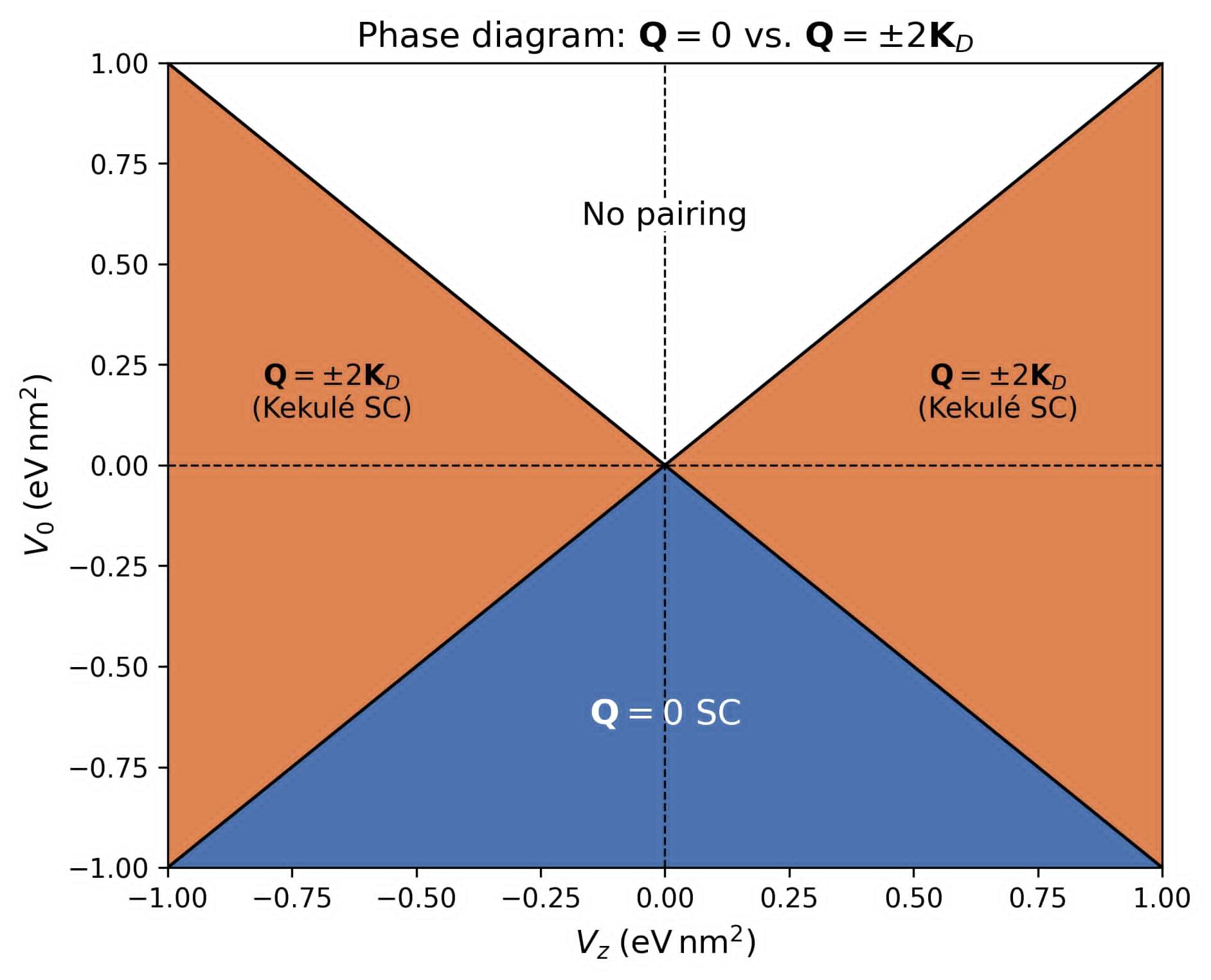}
    \caption{Phase diagram indicating the competition of the inter-valley (${\bm Q}=0$) and intra-valley $({\bm Q} = \pm 2 {\bm K}_D$) superconducting state in $(V_z,V_0)$-interaction phase space.}
    \label{fig:SCphaseinteractions}
\end{figure}

In the purely antisymmetric limit, $V_{1A}=-V_{JB}=-|V_z|$, (i.e. $V_0=0$ and $V_z=-|V_z|$), the ${\bm Q}={\bm 0}$ channel is then completely suppressed as $V_0 =0$, whereas the finite-momentum channel remains active. Since exchanging the labels $A$ and $B$ reverses the sign of $V_z$ and $\kappa$, we choose $V_{1A}=-V_{JB}=-|V_z|$ and $\kappa=1$. The transition temperature is determined by $|V_z|\eta(T_c)=2$. Thus, in the purely antisymmetric limit, the finite-momentum superconducting state with ${\bm Q}=\pm 2{\bm K}_D$ has the highest transition temperature. The phase diagram for the various inter- and intra-valley superconducting states in the phase space of interactions $(V_0,V_z)$ is presented in Fig.~\ref{fig:SCphaseinteractions}. Based on the mean-field analysis, the intra-valley superconducting state exhibits higher transition temperatures for $|V_{z}| > |V_{0}|$, while the inter-valley state requires $|V_{0}| > |V_z|$ and $V_0, V_z <0$. 

\subsection{Superconductivity for $\lambda \neq 0$}

In the presence of a Kane–Mele mass term, $\lambda\neq 0$, the spin-up and spin-down sublattice/layer spinors acquire unequal weights on the $1A$ and $JB$ orbitals. This orbital asymmetry significantly affects the projected interaction matrix elements. The parameter $X=\lambda/\mu$ characterizes this asymmetry and measures the position of the Fermi energy $\mu$ relative to the Kane-Mele gap. The projected interaction matrix elements acquire an explicit $X$ dependence, modifying the interaction kernels in both the ${\bm Q}={\bm 0}$ and ${\bm Q}=\pm 2{\bm K}_D$ sectors as shown in Appendix~\ref{app:AppendixMatelements}. 
 
The contribution of the symmetric component $V_0$ associated to the inter-valley scattering terms ($V_{+-}^{+-},V_{-+}^{-+}$) is enhanced by $1+X^2$, while the orbital-antisymmetric interaction contribution $V_z$ acquires a finite projection proportional to $X$. At the same time, inter-valley exchange scattering terms ($V_{+-}^{-+},V_{-+}^{+-}$) only involve the symmetric component of the interaction $V_0$ and are reduced by $1-X^2$. The linearized gap equation in the ${\bm Q}=0$ becomes,
\begin{equation}
\label{eq:Linearized_gap_Q0_withSOC}
\begin{split}
\begin{pmatrix}
\Delta_{+-} \\
\Delta_{-+}
\end{pmatrix}
&=
-\frac{V_0\eta(T_c)}{2}
\begin{pmatrix}
1+X^2 & 1-X^2 \\
1-X^2 & 1+X^2
\end{pmatrix}
\begin{pmatrix}
\Delta_{+-} \\
\Delta_{-+}
\end{pmatrix}
\\
&\quad
-
V_z X\eta(T_c)
\begin{pmatrix}
1 & 0 \\
0 & -1
\end{pmatrix}
\begin{pmatrix}
\Delta_{+-} \\
\Delta_{-+}
\end{pmatrix}.
\end{split}
\end{equation}

Alternatively, in the finite-momentum sector ${\bm Q}=\pm2{\bm K}_D$, SOC does not change the chiral structure of the interaction kernel. However, the overall strength of both the intra-valley scattering and Cooper pair tunneling interactions is reduced by a factor $1-X^2$. Using the same chiral ansatz as above, the gap equation becomes
\begin{equation}
\label{eq:Linearized_gap_Q2KD_withSOC}
\begin{pmatrix}
\Delta_{--} \\
\Delta_{++}
\end{pmatrix}
=
-\frac{(V_0+\kappa V_z)\eta(T_c)}{4}
(1-X^2)
\begin{pmatrix}
1 & 1 \\
1 & 1
\end{pmatrix}
\begin{pmatrix}
\Delta_{--} \\
\Delta_{++}
\end{pmatrix},
\end{equation}
with transition condition $(V_0+\kappa V_z) (1-X^2) \eta(T_c) =-2$. Therefore, a finite Kane-Mele SOC does not alter its phase-winding structure.

We first consider the purely symmetric limit $V_z=0$. For the ${\bm Q }=0$ channel, the two eigenmodes are $
\vec{\Delta}_{{\bm Q}={\bm 0}}({\bm p})=\Delta(1,\;-1)^T$, with the transition temperature determined by $ V_0X^2\eta(T_c)=-1$, and $\vec{\Delta}_{{\bm Q}={\bm 0}}({\bm p})=\Delta(1,\;1)^T$ with $V_0\eta(T_c)=-1$.
The first is a $\pi$-phase solution, while the second is the uniform inter-valley solution. Since $0<X^2<1$, the effective attraction in the $\pi$-phase channel is always weaker. Therefore, for an attractive symmetric interaction, the uniform ${\bm Q}=0$ state remains the leading instability. The finite-momentum channel instead satisfies $|V|(1-X^2)\eta(T_c)=2,$
for $V_0=-|V|$. This state is always subleading to the uniform inter-valley state, because its effective pairing strength is reduced by the factor $(1-X^2)/2<1$.

We now consider $V_0=0$, i.e the purely antisymmetric limit of the interaction. For the ${\bm Q}=0$ sector, there are two possible inter-valley solutions, $ V_zX\eta(T_c)=-1,$ with $\vec{\Delta}_{{\bm Q}={\bm 0}}({\bm p})=(\Delta,\;0)^T$, and $  V_zX\eta(T_c)=+1$, with $\vec{\Delta}_{{\bm Q}={\bm 0}}({\bm p})=(0,\;\Delta)^T$. Thus, finite SOC activates an inter-valley instability that is absent in the purely antisymmetric channel at $\lambda=0$. This state corresponds to superconductivity on one sublattice/layer, whose physical origin we discuss below. For purely antisymmetric interactions $V_0=0$, the transition temperatures of the inter-valley and intra-valley pairing can be written as
\begin{align}
\label{eq:Tc_inter_log}
&\ln\left(\frac{k_B T_c^{(\mathrm{inter})}}{1.14\,\omega_c}\right)
=
-\frac{1}{|V_z|N(\mu)X},
\qquad \qquad
{\bm Q}={\bm 0},
\\
\label{eq:Tc_intra_log}
&\ln\left(\frac{k_B T_c^{(\mathrm{intra})}}{1.14\,\omega_c}\right)
=
-\frac{2}{|V_z|N(\mu)(1-X^2)},
\quad
{\bm Q}=\pm 2{\bm K}_D .
\end{align}
where $\omega_c$ denotes the pairing interaction window, and the density of states per spin per valley at the Fermi energy can be expressed as,
\begin{equation}
N(\mu)
=
\frac{1}{2\pi J\xi_J^{2/J}}
\frac{\mu}{(\mu^2-\lambda^2)^{(J-1)/J}}
\Theta(\mu-\lambda).
\end{equation}

The leading superconducting instability undergoes a transition at $X = \sqrt{2}-1$. At the band edge, where $\mu=\lambda$ and $X=1$, the inter-valley state, $(\Delta,\; 0)$ or $(0,\; \Delta)$ depending on the sign of $V_z$, has the higher critical temperature. In the opposite limit, $X\to 0$, corresponding to $\mu\gg\lambda$, the intra-valley paired state has the higher critical temperature. The superconducting phase boundary can be determined from $T_c^{(\mathrm{intra})} \ge T_c^{(\mathrm{inter})}$ which gives $\mu \ge (\sqrt{2}+1)\lambda$. Remarkably, this criterion is independent of the chirality index $J$ and of the overall interaction strength $|V_z|$.

\begin{figure}
    \centering
    \includegraphics[width=1\linewidth]{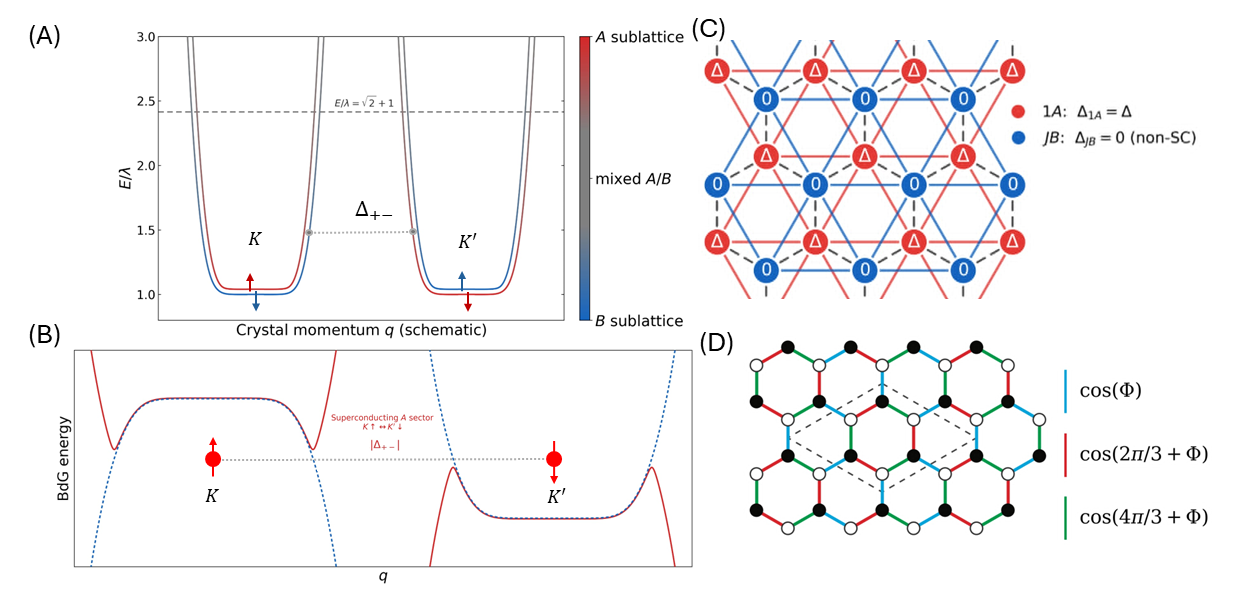}
    \caption{Schematic origin of the inter-valley superconducting state induced by Kane--Mele spin--layer locking near the conduction-band edge. (a) Spin-resolved conduction bands near the valleys ${\bm K}_D$ and $-{\bm K}_D$. Red and blue indicate predominantly outer-layer $1A$ and $JB$ character, respectively, while gray indicates mixed sublattice character. Near the band edge, the $({\bm K}_{D},\uparrow)$ and $(-{\bm K}_{D},\downarrow)$ states are $1A$ polarized and form the inter-valley pair $\Delta_{+-}$. (b) Corresponding BdG space spectra. The $1A$-polarized $({\bm K}_{D},\uparrow; -{\bm K}_{D},\downarrow)$ sector becomes superconducting (due to attraction) and develops a gap $2|\Delta_{+-}|$, whereas the $JB$-polarized 
$({\bm K}_{D},\downarrow; -{\bm K}_{D},\uparrow)$ sector remains normal and gapless. Away from the band edge, increasing sublattice mixing weakens this spin--layer locking and favors the intra-valley pairing channel. 
(c) Even $J$ real-space representation of the sublattice-selective pairing for $V_{1A}<0<V_{JB}$, with $\Delta_{1A}=\Delta$ and $\Delta_{JB}=0$. (d) Kekul\`{e} superconducting pattern defined on the projected honeycomb lattice for even values of $J$, for odd $J$ RMG stacks the projected lattice is a bipartite triangular lattice.}
\label{fig:KMtransition}
\end{figure}

To understand the origin of the inter-valley state at $\lambda\neq 0$, as well as the phase transition in the purely antisymmetric interaction regime, it is useful to consider the limiting cases $X\to 0$ and $X\to 1$. As illustrated in Fig.~\ref{fig:KMtransition} a), the Kane-Mele SOC locks the spin and layer/sublattice degrees of freedom near the conduction-band edge. In the valley ${\bm K}_D$, the band-edge states carry the quantum numbers $(\uparrow,1A;\downarrow,JB)$, whereas in the opposite valley $-{\bm K}_D$, they carry $(\downarrow,1A;\uparrow,JB)$. For purely antisymmetric interactions, either $V_{1A}<0$, which favors pairing between electrons on the top-layer $1A$ sites $(1A,{\bm K}_D,\uparrow;\, 1A,-{\bm K}_D,\downarrow)$, or $V_{JB}<0$, which favors pairing between electrons on the bottom-layer $JB$ sites
$(JB,-{\bm K}_D,\uparrow;\,JB,{\bm K}_D,\downarrow)$. This layer/sublattice superconducting state and its quasiparticle spectrum is shown in Fig ~\ref{fig:KMtransition} b) and c). Near the band edge, the spin-layer locking imposed by the Kane-Mele coupling naturally favors inter-valley pairing, as the opposite spin states are in opposite valleys.

As the Fermi energy increases, the system transitions to an intra-valley superconducting state. The origin of this transition is particularly transparent in the limit $\mu\gg\lambda$, or equivalently $X\to 0$. In this regime, the spin-resolved conduction-band spinors acquire approximately equal weights on the two sublattices or outer layers for $\mu \ge (\sqrt{2}+1)\lambda$, as indicated by the dotted line in Fig.~\ref{fig:KMtransition} a), thereby weakening the spin-layer locking that favors inter-valley pairing. The system consequently approaches the effectively gapless limit discussed above, in which the intra-valley pairing channel becomes dominant.

\subsection{Symmetry classification of Kekul\'{e} Superconductors}

 \begin{table}
    \centering
    \renewcommand{\arraystretch}{1.3}

    \begingroup
    \setlength{\tabcolsep}{3pt}
    \begin{tabular}{|c|c|c|c|}
        \hline
        \textbf{Orbital}
        & \textbf{Valley}
        & \textbf{Spin}
        & \textbf{Symmetry group}
        \\
        \hline

        Odd-$J$ 
        & Triplet
        & Triplet ($m_s=0$)
        & $U(1)\otimes O_V(3)\otimes U_{S}(1)\otimes \mathbb{Z}_3$
        \\

        Odd-$J$ 
        & Singlet
        & Singlet
        & $U(1)\otimes SU_{V}(2) \otimes SU_S(2) \otimes \mathbb{Z}_3$
        \\

        Even-$J$
        & Triplet
        & Singlet
        & $U(1)\otimes O_V(3)\otimes SU_{S}(2) \otimes \mathbb{Z}_3$
        \\

        Even-$J$ 
        & Singlet
        & Triplet ($m_s=0$)
        & $U(1)\otimes SU_{V}(2) \otimes U_S(1)\otimes \mathbb{Z}_3$
        \\
        \hline
    \end{tabular}
    \endgroup
\caption{Summary of the chiral Kekul\'e superconducting states
    for layer-symmetric and layer-antisymmetric interactions satisfying
    $|V_z|>|V_0|$ for $\lambda =0$. The spin and valley classifications follow from the
    fermionic antisymmetry of the orbital pairing of the Cooper pairs. The
    $\mathbb{Z}_3$ symmetry is associated with the commensurate
    center-of-mass momenta $\bm{Q}=\pm 2\bm{K}_D$.}
    \label{tab:Kekule_SC_states}
\end{table}

The proposed superconducting state exhibits a spatial Kekul\'{e} pattern on the projected bipartite lattice, such that the superconducting unit cell is tripled relative to the original lattice unit cell as shown in Fig.~\ref{fig:KMtransition} d)~\cite{fcdc-9lm3,barlas2025quantumgeometryinducedkekule}. The two valley components have equal magnitudes and are related by time reversal. Their opposite chiralities therefore produce a fully gapped, time-reversal-invariant superconducting state. Although each valley component is individually chiral, the combined state preserves time-reversal and inversion symmetry.

Since spin-(valley) correspond to good (approximately good) quantum numbers, the possible Cooper-pair channels can be decomposed into product states associated with spin, valley, and orbital degrees of freedom. One can classify the symmetric properties of the Kekule superconductors for $\lambda =0 $, which depend on the parity of the chiral index $J$. Since the interaction is restricted to opposite-spin pairing, for odd $J$ the chiral Kekul\'{e} superconductor is either a valley-triplet and a spin-triplet ($m_{s}=0$) state or a valley-singlet and a spin-singlet state. In the valley-rotation-symmetric limit, its symmetry structure is characterized by $U(1)\otimes O_V(3)\otimes U_{S}(1) \otimes \mathbb Z_3$ and $U(1)\otimes SU_V(2) \otimes SU_S(2) \otimes \mathbb Z_3$, respectively. Here, $U(1)$ is associated with the overall superconducting phase, $O_V(3)$ denotes the valley-triplet rotational symmetries, and $\mathbb Z_3$ is associated with the lattice translational symmetry.  For even $J$, fermionic antisymmetry permits either a valley-triplet, spin-singlet state, characterized by $U(1)\otimes O_V(3)\otimes SU_{S}(2)\otimes \mathbb Z_3$, or a valley-singlet, spin-triplet ($m_{s}=0$) state, characterized by $U(1)\otimes SU_{V}(2) \otimes U_{S}(1) \otimes \mathbb Z_3$. The symmetry of the Kekul\`{e} superconductor is summarized in Table~\ref{tab:Kekule_SC_states}. In the presence of a finite Kane-Mele spin-orbit coupling, the spin singlet-rotation symmetry is reduced to $SU_S(2) \to U_S(1)$.

\section{Superconducting Phase Diagram and Critical Temperature Scaling}
\label{Section: SC phase}

\begin{figure}
    \centering
    \includegraphics[width=1\linewidth]{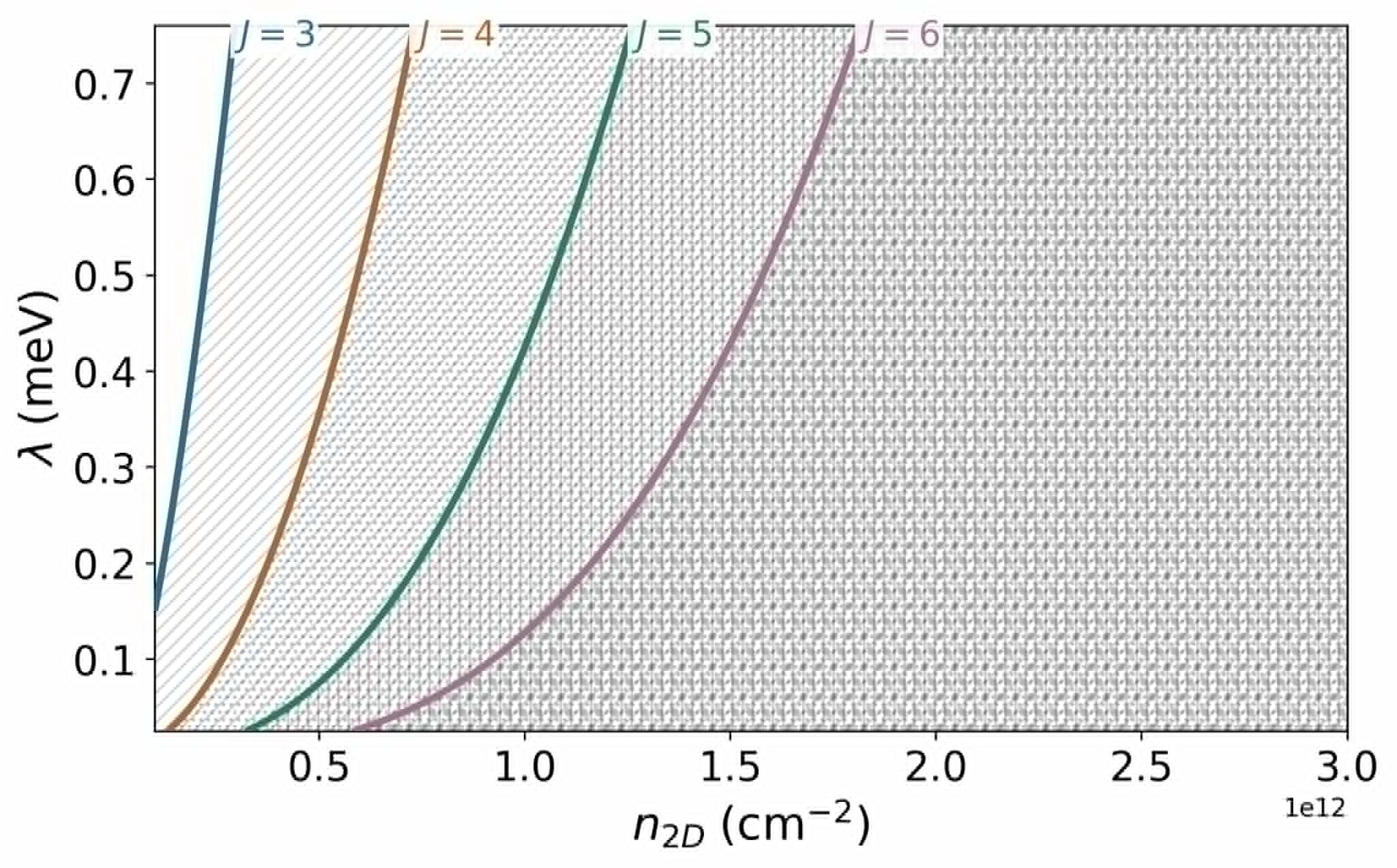}
    \caption{
    Phase boundary between the conventional and Kekul\'e superconducting regimes in the $(n_{2D},\lambda)$ plane. The boundary is determined by $X=\lambda/\mu=\sqrt{2}-1$. The Kekul\'e state is favored at densities above $n^{\mathrm{crit}}_{K}(\lambda,J)$.
    }
    \label{fig:phase_boundary_kekule}
\end{figure}

Now that we have established the existence of intra-valley pairing for $|V_z| \neq 0$, $ |V_{0}|=0$, and $\lambda \neq 0$, we study the superconducting phase diagram and critical temperature scaling as a function of the chirality index $J$. In experiments, a natural tuning parameter is the carrier density $n_{2D}$ rather than $X$. For a chiral dispersion of order $J$, the transition criterion $\mu \geq (\sqrt{2}+1) \lambda$ corresponds to the critical density
\begin{equation}
\label{eq:ncrit}
n^{\mathrm{crit}}_K(\lambda,J)
=
\frac{g}{4\pi}
\left[
\frac{2(1+\sqrt{2})\lambda^2}{\xi_J^2}
\right]^{1/J},
\end{equation}
where $g$ is the spin valley degeneracy.  
The Kekul\'e superconducting order parameter wins out at high enough densities, $n_{2D}\ge n^{\mathrm{crit}}_{K}(\lambda,J).$ Thus, within the minimal model, the system must be doped sufficiently above the SOC gap so that $X<\sqrt{2}-1$. Figure~\ref{fig:phase_boundary_kekule} summarizes this criterion as a phase boundary in the $(n_{2D},\,\lambda)$- plane for several chiralities $J$. Increasing either $J$ or $\lambda$ raises $n^{\mathrm{crit}}_{K}$, meaning that thicker stacks or stronger SOC require higher doping to enter the Kekul\'e regime.

Within the present model, $T_c$ depends on the chirality $J$, carrier density $n_{2D}$, Kane--Mele SOC strength $\lambda$, and effective attractive interaction $|V_z|$. We characterize the interaction regime using the dimensionless BCS pairing strength $w(\mu)\equiv |V_z|N(\mu)$, where $N(\mu)$ is the density of states at the chemical potential. In the numerical estimates below, we set the weak-coupling limit as $w(\mu)\lesssim 0.5$. This value should not be interpreted as a sharp boundary between weak- and strong-coupling regimes, but rather as a practical criterion for identifying a regime in which the weak-coupling BCS treatment remains qualitatively reliable. Larger values of $w(\mu)$ require strong-coupling approaches such as Eliashberg theory, which are beyond the scope of this paper.

\begin{figure}
\centering
\includegraphics[width=\linewidth]{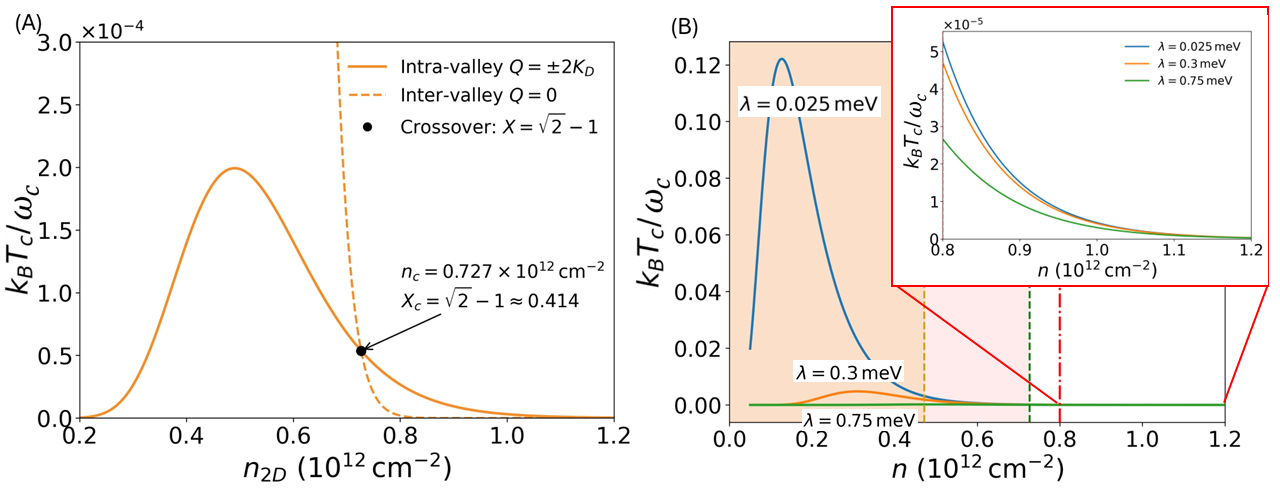}
\caption{
(a) Dimensionless transition temperatures of the conventional and Kekul\'e superconducting channels for $J=4$, $|V_z|=0.4~\mathrm{eV\,nm^2}$, and $\lambda=0.75~\mathrm{meV}$. The crossing marks the density-driven transition into the Kekul\`{e}-dominated regime.
(b) Kekul\`{e} transition temperature as a function of carrier density for several Kane--Mele SOC strengths. The dashed lines mark the phase-boundary condition (green), the operational weak-coupling boundary (orange), and the density cutoff below which trigonal-warping effects (red) cannot be neglected.}
\label{fig:Tc_crossover_conventional_kekule}
\end{figure}

Fig~\ref{fig:Tc_crossover_conventional_kekule}(a) compares the conventional and Kekul\'e transition temperatures obtained from Eqs.~\eqref{eq:Tc_inter_log} and \eqref{eq:Tc_intra_log}. The horizontal axis denotes the carrier density $n_{2D}$, while the vertical axis shows the dimensionless transition temperature
$\bar{T}_c\equiv k_B T_c/\omega_c$. At low density, the conventional inter-valley channel has the larger transition temperature, whereas the Kekul\'e channel is strongly suppressed. As $n_{2D}$ increases, the conventional transition temperature decreases rapidly. By contrast, the Kekul\`{e} transition temperature initially increases, reaches a maximum at $X=\sqrt{1-2/J} $, and then decreases at higher density. The two transition temperatures become equal at the critical density determined by $X=\sqrt{2}-1$. Thus, the density-driven crossover into the Kekul\'e regime can occur while the transition temperature remains finite.

Fig~\ref{fig:Tc_crossover_conventional_kekule}(b) shows the Kekul\'e transition temperature for several values of the Kane--Mele SOC strength. We use the experimentally determined range of the Kane--Mele SOC coupling $\lambda\in [0.025,\, 0.75]~\mathrm{meV}$~\cite{Arp2024,Yang2025,Patterson2025}. To define a common conservative density window for all curves, both the phase and weak-coupling boundaries are evaluated using the largest SOC strength, $\lambda=0.75~\mathrm{meV}$. The green dashed line marks the condition $X=\sqrt{2}-1$; densities to its left are excluded because the Kekul\`{e} channel is not yet the leading instability for the reference SOC strength. The orange dashed line marks the condition $w(\mu)=0.5$; results to its left lie outside our adopted weak-coupling regime and should not be interpreted quantitatively. Furthermore, we restrict our quantitative analysis to $ n_{2D}\geq 0.8\times10^{12}~\mathrm{cm}^{-2}$,
where the Fermi surface is expected to be approximately circular and simply connected ~\cite{barlas2025quantumgeometryinducedkekule,fcdc-9lm3,Han2025}. The red dashed line marks this trigonal-warping cutoff; the minimal model should be reliable at higher densities.


Figure~\ref{fig:Tc_crossover_conventional_kekule}(b) inset focuses on those relevant density interval
$ n_{2D}\in[0.8,1.2]\times10^{12}~\mathrm{cm}^{-2}$. For the reference value $\lambda=0.75~\mathrm{meV}$, the weak-coupling condition requires $ n_{2D}\gtrsim 0.47\times10^{12}~\mathrm{cm}^{-2}$ with $|V_z|\simeq0.4~\mathrm{eV\,nm^2}$, while the condition that the Kekul\`{e} channel be dominant requires $ n_{2D}\gtrsim 0.72\times10^{12}~\mathrm{cm}^{-2}$. Consequently, the trigonal-warping cutoff at
$n_{2D}=0.8\times10^{12}~\mathrm{cm}^{-2}$ is the most restrictive of the three conditions, and all three are satisfied throughout the displayed density window.

Within this window, the Kekul\'e transition temperature decreases monotonically with increasing carrier density, while its dependence on $\lambda$ becomes relatively weak. This behavior follows from Eq.~\eqref{eq:Tc_intra_log}. As the density increases, $\mu$ increases and $X=\lambda/\mu\rightarrow0$, giving $\bar{T}_c \simeq 1.14 \exp\left[-2/(|V_z|N(\mu))\right],$
with $N(\mu)\simeq (2\mu^{2/J-1})/(\pi J\xi_J^{2/J})$. The explicit SOC dependence is therefore suppressed in the higher-density limit, and the transition temperature is controlled primarily by $\mu$, $J$, and $|V_z|$, with $\mu$ determined by $n_{2D}$. For rhombohedral tetralayer graphene $(J=4)$ in the parameter regime considered here, $ \bar{T}_c\sim1\times10^{-5}
\text{--}5\times10^{-5}$. Taking $\omega_c=0.2~\mathrm{eV}$, this corresponds to $ T_c\sim23\text{--}116~\mathrm{mK}$ for $|V_z|\simeq0.4~\mathrm{eV\,nm^2}$ chosen phenomenologically~\cite{PhysRevB.105.L100503}. For instance, if the interaction is doubled to $|V_z| \simeq 0.8~\mathrm {eV \,nm^2}$, then $T_c$ in the regime could reach $2.3-18.5 ~\mathrm{K}$. 

In the low-to-intermediate density range, increasing $J$ can increase $T_c$; however, at sufficiently high densities, the trend reverses. This behavior follows directly from Eq.~\eqref{eq:Tc_intra_log}. Given $ \bar{T}_{c,J} =1.14\,e^{-g_J}, $
substituting the chiral-$J$ density of states into Eq.~\eqref{eq:Tc_intra_log} gives
\begin{equation}
    g_J(n_{2D})
    =
    \frac{4\pi J}{|V_z|}
    \frac{\mu_J(n_{2D})}{k_F^2}.
    \label{eq:wJ_density}
\end{equation}
At fixed carrier density, $k_F$ is fixed, and the prefactor
$(4\pi)/(|V_z|k_F^2)$
is independent of the chirality. Therefore, the $J$ dependence of $T_c$ is controlled entirely by the product
$J\mu_J(n_{2D}).$ At intermediate densities, increasing $J$ flattens the chiral dispersion and can reduce $\mu_J$ sufficiently to enhance $T_c$. At higher density, $X_J \ll 1,$ where the chemical potential approaches
$\mu_J\simeq \xi_Jk_F^J$, we obtain $J\mu_J \simeq J\gamma_1\left((\hbar v_Fk_F/\gamma_1) \right)^J$. Consequently, the high-density inversion of $T_{\mathrm{c}} $ between two different chiralities is determined by the band dispersion and independent of the interaction strength. In practice, this inversion occurs at high densities $n_{2D} \sim 7 \times 10^{12} \mathrm{cm}^{-2}$ where our estimates of $|V_{z}|$ yield very low values for $T_c$.


\section{Superfluid stiffness}
\label{Section: Superfluid stiffness}

The superfluid stiffness tensor measures the phase rigidity of a superconducting state. For a projected band, it can be decomposed into conventional and geometric contributions, $ D_{\mu\nu} = D^{\mathrm{conv}}_{\mu\nu} +D^{\mathrm{geom}}_{\mu\nu} $ \cite{PhysRevB.95.024515,PhysRevLett.131.016002}. In the weak-coupling regime, the geometric contribution behaves as $ D_s^{\mathrm{geom}}\sim |\Delta|^2/\mu$, therefore it can be neglected when compared to the   
 conventional contribution $ D_s^{\mathrm{conv}}\sim J\mu $ ~\cite{fw3r-pcw5,barlas2025quantumgeometryinducedkekule}. Because the chiral dispersion is azimuthally symmetric, the conventional stiffness is isotropic,
$ D^{\mathrm{conv}}_{\mu\nu} =D_{s}\delta_{\mu\nu}$. Evaluating the conventional superfluid stiffness, we get
\begin{gather}
D_s
=
\frac{gJ\mu}{\pi}
I(\tilde{\Delta},X),
\label{eq:Dfinite}
\\
I(\tilde{\Delta},X)
=
\int_X^\infty du
\left(
1+\frac{X^2}{u^2}
\right)
\left[
1-
\frac{u-1}{\sqrt{(u-1)^2+\tilde{\Delta}^2}}
\right],
\label{eq:dimensionlessI}
\end{gather}
where $\tilde{\Delta}=\Delta/\mu$. Equation~\eqref{eq:Dfinite} shows that the chirality index $J$ enters explicitly through the prefactor $J\mu/\pi$, while the nontrivial SOC dependence is contained in the dimensionless integral $I(\tilde{\Delta}, X)$. The factor $\mu$ increases with increasing SOC at fixed density, but the lower limit and kernel of Eq.~\eqref{eq:dimensionlessI} reduce the available phase space for $X>0$. Therefore, the net SOC dependence of $D^{\mathrm{conv}}_{\mu\nu}$ must generally be evaluated numerically.

\begin{figure}
\includegraphics[width=.5\textwidth]{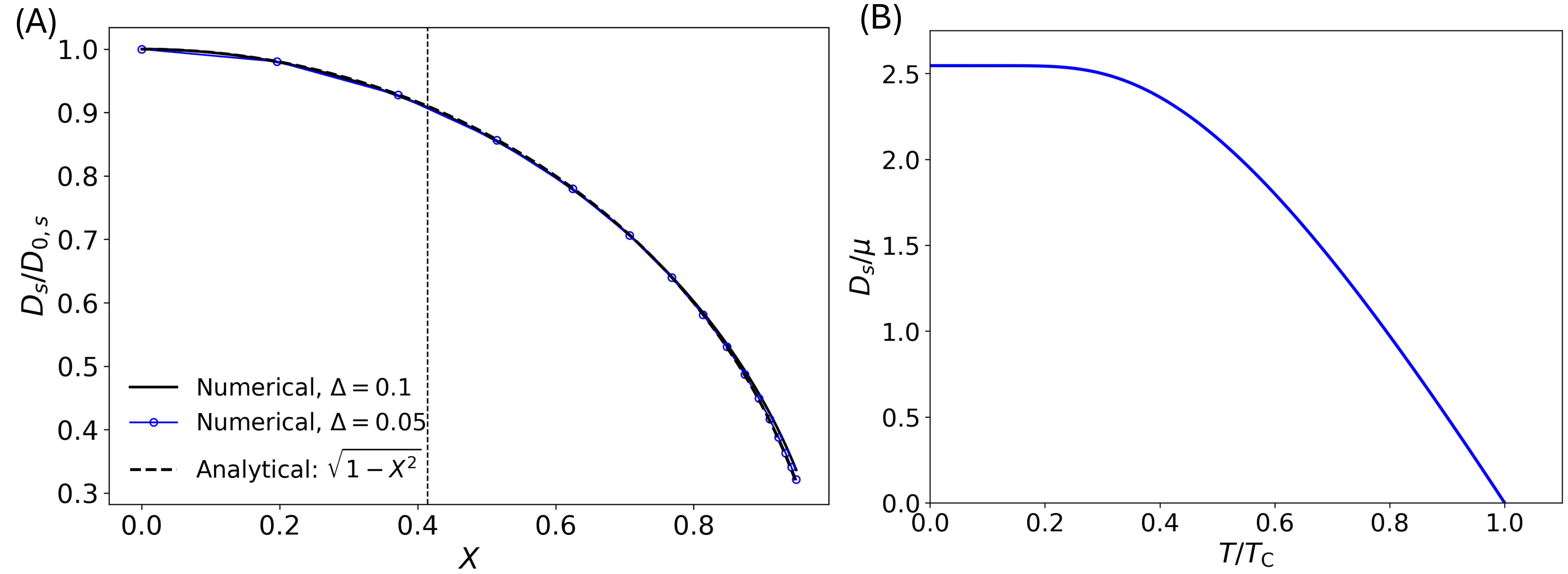}
  \caption{
  Superfluid stiffness of the Kekul\'e superconducting state.
  (a) Normalized stiffness as a function of $X=\lambda/\mu$.
  The stiffness is suppressed as $X$ increases.
  (b) Temperature dependence of the total stiffness for $J=4$.}
  \label{fig:stiffness}
\end{figure}

The integral has useful limiting forms. In the weak-coupling limit, $\tilde{\Delta}\rightarrow0^+$, and in the zero SOC limit, $X\rightarrow0^+$, one finds
\begin{gather}
\lim_{\tilde{\Delta}\rightarrow 0^+}
I(\tilde{\Delta},X)
=
2(1-X^2),
\\
\lim_{X\rightarrow0^+}
I(\tilde{\Delta},X)
=
1+\sqrt{1+\tilde{\Delta}^2}.
\end{gather}
For $X=0$, the stiffness is consistent with previous calculations~\cite{PhysRevB.95.024515,PhysRevLett.131.016002}, and has a weak dependence on $\Delta$ in the weak coupling limit $\tilde{\Delta} \ll 1$. It is useful to compare the finite-SOC stiffness with the $\lambda=0$ result $D_{0,s} $. In the weak-coupling limit, this ratio reduces to $D_{s}(X)/D_{0,s}\sim \sqrt{1-X^2}$, which is independent of the chirality index $J$. Thus, relative to its zero-SOC value, the conventional stiffness is suppressed by a universal factor controlled only by $X=\lambda/\mu$. Fig.~\ref{fig:stiffness} (a) shows this normalized conventional stiffness as a function of $X$. For fixed chemical potential $\mu$, increasing the Kane--Mele SOC strength $\lambda$ increases $X$, so moving from left to right corresponds to increasing SOC. The monotonic decrease in stiffness indicates that SOC suppresses the conventional contribution to phase rigidity. Conversely, for fixed $\lambda$, increasing $\mu$ decreases $X$. Moving from right to left therefore corresponds to increasing density, which enhances the conventional stiffness and stabilizes the Kekul\'e condensate. The temperature dependence of the stiffness, shown in Fig~\ref{fig:stiffness} b), follows the standard temperature scaling.

\begin{figure*}
    \centering
    \includegraphics[width=1\linewidth]{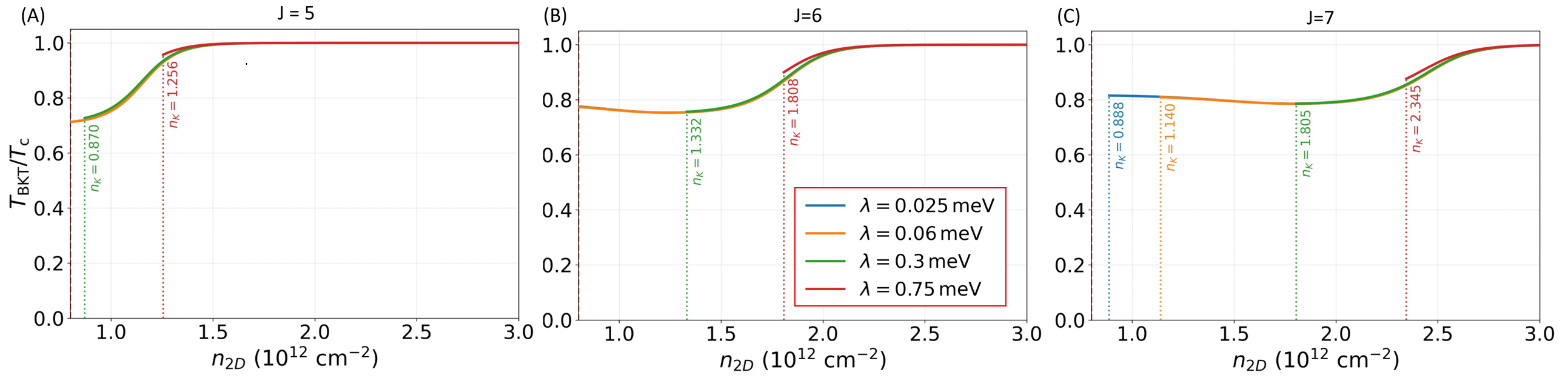}
    \caption{
   Ratio $T_{\mathrm{BKT}}/T_{c}$ as a function of carrier density for (a) $J=5$, (b) $J=6$, and (c) $J=7$ shown for several SOC strengths $\lambda$. The dotted vertical lines indicate the corresponding Kekul\'e-transition densities. The densities are above the trigonal warping scale $n_{2D} \geq 0.8 \times 10^{12}$ cm$^{-2}$. }
    \label{fig:BKT_T}
\end{figure*}

For an isotropic two-dimensional superconductor, the transition is determined by the Nelson--Kosterlitz universal jump condition: $k_B T_{\mathrm{BKT}}=(\pi/8)D_s(T_{\mathrm{BKT}})$~\cite{Kosterlitz1973, PhysRevLett.39.1201,Berezinskii:1970pzv}.  Figure~\ref{fig:BKT_T} a)-c) shows the ratio 
$T_{\mathrm{BKT}}/T_c$ as a function of the carrier density
$n_{2D}$ for several values of the spin-orbit coupling $\lambda$ and $J=5$, $J=6$, and $J=7$, respectively.
The colored dotted vertical lines correspond to critical densities $n_K(\lambda,J)$ associated with the Kekul\'e transition at $ X=\sqrt{2}-1$. Figure~\ref{fig:BKT_T} a)-c) indicate that, for all cases (different values of $J$), at high densities pair formation and phase coherence occur at nearly the same temperature, whereas at low densities
$T_{\mathrm{BKT}}/T_c<1$ signals a phase-fluctuation regime
in which Cooper pairs form at $T_c$ but establish
long-range phase coherence at the lower temperature
$T_{\mathrm{BKT}}$.

The behavior in Fig~\ref{fig:BKT_T} can be understood from the scaling of the superfluid stiffness $D_{s} \propto n_{2D}^{J/2}$ and the critical temperature ($T_c $) scaling as a function of the density and density of states, respectively. At high densities, even though the superfluid stiffness grows as $n_{2D}^{J/2}$, $T_{\mathrm{BKT}}$ is limited by $T_c$, which gets saturated at higher densities. Alternatively, at low densities, while $T_c$ is enhanced exponentially due to the large density of states, the superfluid stiffness reduces, giving a saturation  $T_{\mathrm{BKT}}/T_c\simeq0.8$. Since the superfluid stiffness is weakly dependent on the Kane-Mele SOC, different values of $\lambda$ track each other, overlapping over most of the displayed density range.

\section{Discussion and Outlook}
\label{Section: Discussion}

The intravalley nature of the pairing provides several complementary experimental signatures. Because the two condensates carry center-of-mass momenta ($\pm 2 {\bm K}_{D}$), superconducting Josephson STM should reveal Kekul\`{e}-scale Fourier components at these momenta. Since valley is approximately conserved for smooth interfaces with negligible intervalley scattering, a valley-resolved Andreev spectroscopy experiment can directly distinguish the two cases. Finally, the opposite finite momenta of the two condensates imply a uniform composite charge-($4e$) order parameter, $\Phi_{4e}\sim \Delta_{2 {\bm K}_D}\Delta_{-2 {\bm K}_D}$~\cite{PDWreview}, which can generate an enhanced second harmonic in the Josephson current, $I(\phi)\sim I_2\sin 2\phi$, and, in a regime where the ordinary charge-($2e$) Josephson channel is suppressed, along with signatures associated with ($4e$) transport such as a doubled ac Josephson frequency or ($h/4e$)-scale interference. 

Apart from the superconducting Kekul\`{e} pattern
(see Fig.~\ref{fig:KMtransition}d), the most striking property of the
time-reversal-symmetric Kekul\`{e} state is the presence of two condensates
with opposite orbital chirality. The internal form factors of the two
condensates carry opposite phase windings,
$\Delta_{++}\sim e^{-i J\phi_{{\bm p}}}$ and
$\Delta_{--}\sim e^{iJ\phi_{{\bm p}}}$. Because the underlying RMG lattice contains a $C_{3z}$ rotational symmetry, the internal crystal angular momentum of these chiral states is defined modulo three. One possible way to detect the chirality is through optical excitation. Structured light such as a Laguerre-Gaussian beam~\cite{k9m4-h474,dqv7-w2w4}, carrying orbital angular momentum $\ell$ in addition to photon helicity $\sigma=\pm1$, can be tuned to frequencies just above the superconducting pair-breaking threshold, $\hbar\omega\gtrsim2\Delta$, where the BdG quasiparticle matrix elements retain the chirality of the superconducting order parameter. This can break pairs in one valley condensate, producing a transient imbalance in the quasiparticle populations and superconducting gap amplitudes of the two valleys. The resulting uncompensated orbital chirality can be detected in a pump-probe experiment through a transient Kerr rotation, anomalous Hall response, or circular dichroism. Reversing the angular momentum of optical excitation should excite the opposite chiral sector and reverse the sign of a chirality-sensitive transient response, providing a direct experimental signature of the valley-locked orbital chirality.

Before we conclude, we would like to address the role of remote-hopping
effects neglected in our mean-field calculations. Although we work at
densities above the Lifshitz transition, where the Fermi surface remains
connected, remote hopping breaks azimuthal symmetry through trigonal
warping. In the intra-valley channel, this spoils the perfect nesting
between states at opposite momenta on the same valley Fermi surface and
introduces an energy scale $E_{\mathrm{tri}}$ that cuts off the
weak-coupling Cooper logarithm. Consequently, unlike in the azimuthally symmetric model, a finite critical
pairing strength $|V_z|$ will be required to stabilize Kekul\'{e} superconductivity. 

Finally, a layer-selective proximity effect need not directly realize the idealized
limit $V_A=-V_B$. Layer-dependent Coulomb and screening
corrections can renormalize the two interactions differently and drive the
system into the regime $|V_z|>|V_0|$, in which the finite-momentum
Kekul\`{e} superconducting state is favored. Thus, the phase diagram in Fig.~\ref{fig:SCphaseinteractions} obtained
in terms of $(V_0,V_z)$ applies more generally to strongly layer-asymmetric
effective interactions and does not require the fine-tuned microscopic
condition $V_A=-V_B$.

In summary, we showed that the chiral bands in rhombohedral graphene can naturally support finite-momentum Kekul\'e superconductivity when the microscopic attraction is layer/sublattice antisymmetric, \(V_A=-V_B\). For the chiral Kane--Mele bands, the antisymmetric interaction favors the Kekul\'e state above a universal threshold \(X=\lambda/\mu<\sqrt{2}-1\), corresponding to carrier density $n_{2D} \sim 10^{12} \mathrm{cm}^{-2}$ for experimentally reported values of Kane-Mele SOC $\lambda$. We found that the resulting critical temperature range is $T_{\mathrm{c}} \sim 10 \mathrm{K} - 100 \mathrm{mK}$, and the superfluid stiffness remains positive and dominated by the conventional contribution in the Kekul\`{e} regime. These results demonstrate that band geometry and sublattice-selective interactions can stabilize a finite-\(\bm Q\) superconducting state in graphene-based chiral systems.

\section{Acknowledgment}
The authors acknowledge support from the DOE EPSCoR program under the award DE-SC0022178.

\appendix

\section{Calculation of Interaction matrix elements}
\label{app:AppendixMatelements}

In this Appendix, we calculate the interaction matrix elements defined in Eq.~\ref{eq:projectedV} after projection onto the conduction-band Fermi surface. Owing to the nontrivial wavefunction geometry of the chiral bands in RMG, an interaction that is local in the orbital basis can acquire momentum dependence upon projection onto the conduction band, giving rise to the chiral order parameter discussed in this paper. 

A diagonal microscopic local interaction in the orbital basis, parameterized by $(V_{1A}, V_{JB})$, can be decomposed into an orbital-symmetric component, $V_0=(V_{1A}+V_{JB})/2$, and a purely orbital-antisymmetric component, $V_z=(V_{1A}-V_{JB})/2$. The symmetric and antisymmetric orbital channels are weighted differently in the ${\bm Q}={\bm 0}$ and ${\bm Q}=\pm 2{\bm K}_D$ sectors because the wavefunction weight associated with each orbital component is determined by the band chirality and the Kane-Mele mass term. In the ${\bm Q}={\bm 0}$ sector, the projected interaction for a $\lambda \neq 0 $ is given by,
\begin{equation}
\label{eq:Q0_kernel_finiteSOC}
V_{{\bm Q}={\bm 0}}({\bm p},{\bm p}')
= \frac{V_0}{2}
\begin{pmatrix}
1+X^2 & 1-X^2 \\
1-X^2 & 1+X^2
\end{pmatrix}
+
V_z X
\begin{pmatrix}
1 & 0 \\
0 & -1
\end{pmatrix},
\end{equation}
where $X = \lambda/\mu$. In the gapless case, $\lambda=0$, only the orbital-symmetric interaction survives, as can be seen by setting $X=0$. For finite $\lambda$, the orbital interactions $(V_{1A},V_{JB})$ produce different diagonal and off-diagonal scattering amplitudes in the ${\bm Q}={\bm 0}$ sector. The diagonal component, corresponding to the intervalley scattering interaction, is enhanced by the factor $1+X^2$. For $\lambda\neq 0$, this component also contains a contribution from the orbital-antisymmetric interaction $V_z$, whose projected amplitude is proportional to $X$. Because this contribution has opposite signs in the two valleys, it is repulsive in one valley and attractive in the other. By contrast, the intervalley-exchange interaction, corresponding to the off-diagonal scattering amplitude, contains only the orbital-symmetric component and is reduced by the factor $1-X^2$.

In contrast, the intra-valley sector, ${\bm Q}=\pm 2{\bm K}_D$, contains nontrivial angular dependence. For $\lambda \neq 0$, it can be expressed as,
\begin{equation}
V^{\lambda}_{{\bm Q}=\pm 2{\bm K}_D}({\bm p},{\bm p}')
=
(1-X^2)
V^{\lambda=0}_{{\bm Q}=\pm 2{\bm K}_D}({\bm p},{\bm p}'),
\end{equation}
where
\begin{eqnarray}
V^{\lambda =0}_{{\bm Q}=\pm 2{\bm K}_D}({\bm p},{\bm p}')
&=
\frac{V_0}{2}
\begin{pmatrix}
\cos\delta_{{\bm p},{\bm p}'} & \cos\Sigma_{{\bm p},{\bm p}'} \\
\cos\Sigma_{{\bm p},{\bm p}'} & \cos\delta_{{\bm p},{\bm p}'}
\end{pmatrix}
\\
&\quad
+
\frac{iV_z}{2}
\begin{pmatrix}
\sin\delta_{{\bm p},{\bm p}'} & -\sin\Sigma_{{\bm p},{\bm p}'} \\
\sin\Sigma_{{\bm p},{\bm p}'} & -\sin\delta_{{\bm p},{\bm p}'}
\end{pmatrix},
\end{eqnarray}
and $\delta_{{\bm p},{\bm p}'} = J(\phi_{{\bm p}'}-\phi_{\bm p})$, and $ \Sigma_{{\bm p},{\bm p}'}
=J(\phi_{{\bm p}'}+\phi_{\bm p})$.  For the finite-momentum sector, ${\bm Q}=\pm2{\bm K}_D$, $\lambda \neq 0$ reduces its overall strength by the factor $1-X^2$, but retains the chiral structure of the interaction matrix elements. In both cases, the intra-valley interaction and the Cooper pair tunneling interactions lead to the chiral $J$ superconducting order, as discussed in detail in the main paper.


%

\end{document}